\documentclass[sigconf]{acmart}
\usepackage[utf8]{inputenc}

\PassOptionsToPackage{table,xcdraw}{xcolor} 

\usepackage{textgreek}
\usepackage{multirow}
\usepackage{subcaption}
\usepackage{xcolor,colortbl}
\definecolor{gray}{rgb}{0.1,0.1,0.1}
\usepackage[T1]{fontenc}
\usepackage{graphicx}
\usepackage{tabularx}
\usepackage{longtable}
\usepackage{enumitem}
\usepackage{amsmath}
\usepackage{graphicx}
\usepackage{array}
\usepackage{booktabs}
\usepackage{makecell}
\usepackage{framed} 
\usepackage{ragged2e}
\usepackage{tikz}

\AtBeginDocument{%
  \providecommand\BibTeX{{%
    \normalfont B\kern-0.5em{\scshape i\kern-0.25em b}\kern-0.8em\TeX}}}

\copyrightyear{2026}
\acmYear{2026}
\setcopyright{cc}
\setcctype{by}
\acmConference[CSCW Companion '26]{Companion of the Computer-Supported Cooperative Work and Social Computing}{October 10--14, 2026}{Salt Lake City, UT, USA}
\acmBooktitle{Companion of the Computer-Supported Cooperative Work and Social Computing (CSCW Companion '26), October 10--14, 2026, Salt Lake City, UT, USA}
\acmDOI{10.1145/3785651.3816117}
\acmISBN{979-8-4007-2378-0/2026/10}

\begin{document}
\title[The Capacity to Care: Designing Social Technology for Sustained Engagement With Societal Challenges]{The Capacity to Care: Designing Social Technology for Sustained Engagement With Societal Challenges}

\author{JaeWon Kim}
\orcid{0000-0003-4302-3221}
\affiliation{%
  \institution{University of Washington}
  \city{Seattle}
  \country{USA}}
\email{jaewonk@uw.edu}

\author{Lindsay Popowski}
\orcid{0000-0002-5649-0286}
\affiliation{%
  \institution{Stanford Univeristy}
  \city{Stanford}
  \country{USA}
}
\email{popowski@stanford.edu}

\author{Louisa Conwill}
\authornote{All authors contributed equally to this proposal.}
\affiliation{%
  \institution{University of Notre Dame}
  \city{Notre Dame}
  \country{USA}}
\orcid{0009-0001-7116-266X}
\email{lconwill@nd.edu}

\author{Elizabeth `Lizzie' Li}
\authornotemark[1]
\affiliation{%
  \institution{Northwestern University}
  \city{Illinois}
  \country{USA}}
\orcid{0009-0009-4654-2634}
\email{elizabethli2028@u.northwestern.edu}

\author{Meryl Ye}
\authornotemark[1]
\affiliation{%
  \institution{Carnegie Mellon University}
  \city{Pittsburgh}
  \country{USA}}
\orcid{0000-0001-8215-9020}
\email{merylye@cmu.edu}

\author{Jiaying `Lizzy' Liu}
\affiliation{%
  \institution{Nanyang Technological University}
  \city{Singapore}
  \country{Singapore}}
\orcid{0000-0002-5398-1485}
\email{jiayingliu@utexas.edu}

\author{Jose A. Guridi}
\affiliation{%
  \institution{Cornell University}
  \city{Ithaca}
  \country{USA}}
\orcid{0000-0003-0543-699X}
\email{jg2222@cornell.edu}

\author{Theia Henderson}
\affiliation{%
  \institution{MIT CSAIL}
  \city{Cambridge, MA}
  \country{USA}}
\orcid{0000-0002-0263-9786}
\email{theia@mit.edu}

\author{Bingxu Han}
\affiliation{%
  \institution{Stanford University}
  \city{Stanford}
  \country{USA}}
\orcid{0009-0002-3403-6338}
\email{bingxu9@stanford.edu}

\author{Dennis Wang}
\affiliation{%
  \institution{National Yang Ming Chiao Tung University}
  \country{Taiwan}}
\orcid{0000-0000-0000-0000}
\email{dennis@nycu.edu.tw}

\author{Angel Hsing-Chi Hwang}
\affiliation{%
  \institution{University of Southern California}
  \city{Los Angeles}
  \country{United States}}
\orcid{0000-0002-0951-7845}
\email{angel.hwang@usc.edu}

\author{Susan Wyche}
\affiliation{%
  \institution{Michigan State University}
  \country{USA}}
\orcid{0000-0000-0000-0000}
\email{spwyche@msu.edu}

\author{Yasmine Kotturi}
\affiliation{%
  \institution{University of Maryland, Baltimore County}
  \city{Baltimore}
  \country{USA}}
\orcid{0000-0001-6201-7922}
\email{kotturi@umbc.edu}

\author{Gillian R. Hayes}
\affiliation{%
  \institution{University of California, Irvine}
  \city{Irvine}
  \country{USA}}
\orcid{0000-0003-0966-8739}
\email{gillianrh@ics.uci.edu}

\author{Angela D. R. Smith}
\affiliation{%
  \institution{University of Texas at Austin}
  \city{Austin}
  \country{USA}}
\orcid{0000-0001-5546-5452}
\email{adrsmith@utexas.edu}

\renewcommand{\shortauthors}{JaeWon Kim, et al.}

\begin{abstract}
    People care about climate change, injustice, and humanitarian crises. The challenge is not apathy but capacity: sustained engagement with large-scale problems is psychologically costly, and social media architecture often amplifies awareness while providing few pathways to meaningful action. The result is rising distress, overwhelm, and disengagement---particularly among young people who encounter global suffering through platforms designed for attention capture rather than constructive response. This workshop examines how social media design shapes the conditions for sustained engagement with societal challenges. Drawing on Tronto's care ethics framework and research in moral psychology and platform studies, we ask why caring at scale is difficult and how social media can both exacerbate and potentially mitigate this difficulty. Tronto's framework shows that good care requires more than awareness: it demands responsibility, competence, and community. Dominant social media architectures stall the caring process at its earliest phase. We invite researchers and designers to identify platform designs that deplete or support the capacity to care, and to develop design directions for \textit{sustainable care}: engagement that people can maintain over time without burning out.
\end{abstract}



\maketitle

\section{Introduction}
\label{sec:introduction}
Care ethics offers a framework for understanding why social media often widens the gap between awareness and action on collective problems. Joan Tronto's account of ``good care'' describes a process that moves through distinct phases, from noticing a need to responding to it within a community. This proposal argues that current platform architectures systematically disrupt that process. We call the alternative design goal \textit{sustainable care}~\cite{kim2026sustainable}: productive engagement with societal challenges that people can maintain over time, because the sociotechnical environment supports the full caring process rather than undermining it.

\subsection{Background: What Does Good Care Look Like?}
Care ethics, a tradition grounded in feminist moral philosophy~\cite{held2006ethics}, treats care as comprehensive and specific. Fisher and Tronto define it as ``a species \textit{activity} that includes everything that we do to maintain, continue, and repair our `world' so that we can live in it as well as possible.''~\cite{tronto2013caring, fisher1990toward}. \textit{Care is not just a sentiment; it is a practice.}

\begin{figure}[h]
\centering
\begin{tikzpicture}[
  phase/.style={draw, rounded corners=6pt, minimum width=2.8cm, minimum height=1.1cm, align=center, font=\small},
  arrow/.style={->, >=stealth, thick},
  every node/.style={font=\small}
]

\node[phase] (about) at (0, 2.4) {\textbf{1. Caring About}\\[1pt]\textit{Attentiveness}};
\node[phase] (for) at (4.2, 2.4) {\textbf{2. Caring For}\\[1pt]\textit{Responsibility}};
\node[phase] (giving) at (4.2, 0) {\textbf{3. Caregiving}\\[1pt]\textit{Competence}};
\node[phase] (receiving) at (0, 0) {\textbf{4. Care Receiving}\\[1pt]\textit{Responsiveness}};

\draw[arrow] (about) -- (for);
\draw[arrow] (for) -- (giving);
\draw[arrow] (giving) -- (receiving);
\draw[arrow] (receiving) -- (about);

\draw[dashed, rounded corners=12pt, gray] (-1.8, -0.9) rectangle (6.0, 4.3);
\node[font=\small, gray, align=center] at (2.1, 3.7) {\textbf{5. Caring With}\\[1pt]\textit{Solidarity: plurality, communication, trust, respect}};

\end{tikzpicture}
\caption{Tronto's phases of care. Good care is a cyclical process: each phase carries a moral requirement, and the response to care given generates renewed attentiveness. The fifth phase, caring with, requires that all parties can participate in defining needs and deciding how to meet them~\cite{tronto2013caring, sevenhuijsen1998citizenship}.}
\label{fig:care-process}
\end{figure}
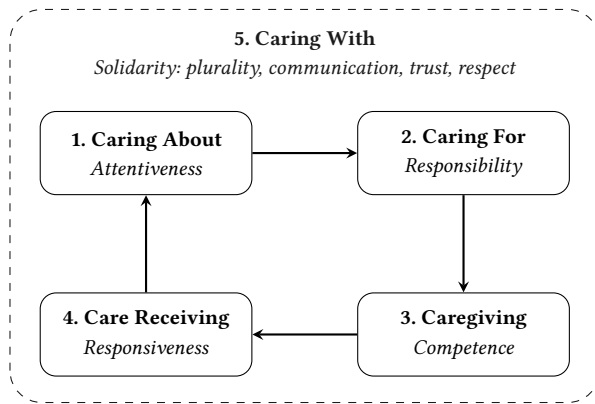

Tronto identifies four phases of this practice, each with a corresponding moral dimension~\cite{tronto1993moral, tronto2013caring}. \textit{Caring about} means noticing that a need exists; its moral quality is \textit{attentiveness}. \textit{Caring for} means taking responsibility for meeting that need; its moral quality is \textit{responsibility}. \textit{Caregiving} is the actual work of responding; its moral quality is \textit{competence}. \textit{Care receiving} requires attending to whether the care was adequate and what new needs have emerged; its moral quality is \textit{responsiveness}. Because each response generates renewed attentiveness, the process is cyclical (Figure\ref{fig:care-process}).

In later work, Tronto introduces a fifth phase, \textit{caring with}. \textit{Caring with} requires that everyone involved, givers and receivers alike, can participate in deciding what care is needed and how to provide it~\cite{tronto2013caring}. Rather than representing another step in the care cycle, \textit{caring with} situates care within the broader social and institutional conditions that make it possible.

This framework clarifies what ``good care'' actually requires. Awareness alone does not constitute care; neither does good intention. Moving from noticing a need to meeting it takes judgment, capacity, knowledge, and community (Table~\ref{tab:care-phases}). Care has never been easy to sustain; Tronto's framework exists in part because care has long been undervalued and poorly distributed~\cite{tronto1993moral}. What has changed is the sociotechnical environment in which people encounter these needs: always-on, personalized social media, accessed via ubiquitous devices and structured by attention-driven platform designs. As the next section argues, dominant platform architectures amplify attentiveness while undercutting the conditions that would make it bearable.

\begin{table*}[h]
\centering
\caption{Tronto's phases of care applied to social media. Each phase requires a moral quality that current platforms tend to undermine but could support through different design choices.}
\label{tab:care-phases}
\small
\begin{tabular}{p{3cm} p{4.8cm} p{4.8cm}}
\toprule
\textbf{Phase (Moral Quality)} & \textbf{Social Media Status Quo} & \textbf{Social Media Design Potential} \\
\midrule
1. Caring About \newline (Attentiveness)
  & Abundant exposure to suffering with no natural stopping points; stepping away feels like abandonment
  & Bounded awareness that respects capacity; natural endpoints for engagement \\
\addlinespace
2. Caring For \newline (Responsibility)
  & Broadcast structure leaves responsibility diffuse; everyone sees the same content, but no one is positioned to act
  & Connecting users to communities where responsibility can be shared and made specific \\
\addlinespace
3. Caregiving \newline (Competence)
  & Emotional framing over actionable knowledge; few pathways from awareness to material response
  & Surfacing what people can actually do; supporting organized, collective response \\
\addlinespace
4. Care Receiving \newline (Responsiveness)
  & Negative stories dominate; little visibility of whether care efforts are working or what needs remain
  & Making collective efforts and their effects visible; surfacing emerging needs \\
\addlinespace
5. Caring With \newline (Solidarity)
  & Isolated individuals witnessing collective problems; weak sense of shared membership
  & Community structures where members share in defining needs and deciding how to respond \\
\bottomrule
\end{tabular}
\end{table*}

\subsection{Problem Area: Why Good Care Becomes Unsustainable}
Public concern about global issues like climate change, political injustice, and humanitarian crises has grown substantially in recent years~\cite{Courtney2025ReportZi, leiserowitz2023climate, Briggs2024GenPresentc, UnitedWayNCA2024GenSurveyq}. Yet this rising awareness has not translated straightforwardly into sustained action. Instead, it frequently co-occurs with distress, overwhelm, and disengagement. A 2025 GlobeScan study found that 38\% of Gen Z report feeling stressed or anxious about climate change most or all of the time, and 17\% of older generations reporting the same~\cite{Courtney2025ReportZi}. Clayton~\cite{clayton2020climate} found that, with respect to climate anxiety, the combination of perceived threat and perceived helplessness are observed with withdrawal rather than action. Ford et al.~\cite{ford2023political} find that daily exposure to political events simultaneously increases motivation to act and worsens emotional well-being.

This pattern is precisely what Tronto's framework would predict when care stalls at its first phase. Singer and Klimecki~\cite{singer2014empathy} distinguish between empathic distress, feeling others' pain as one's own, which leads to withdrawal, and compassion, an other-oriented response that sustains action. Which response takes hold depends not on how much someone cares, but on whether they have the support and pathways to respond constructively. Attentiveness without the conditions to move further through the caring process---without distributed responsibility, competence, and community---collapses inward. The result is burnout, not because people cared too little, but because they cared without the support to do so well~\cite{tronto2013caring}.

Research on civic engagement bears this out. Flanagan and Bundick~\cite{flanagan2011civic} find that collective action correlates with hope, efficacy, and meaning, but that intense engagement without social support carries psychosocial costs. Gilster~\cite{gilster2012comparing} finds that neighborhood-level activism builds empowerment and local social ties, suggesting that benefits erode when the caring burden is not distributed. In each case, sustainable care depends on structural conditions, not just individual willpower.

\subsection{Social Media Design and Sustainable Care}
Social media plays a dual role in the problem of sustainable care. Its architecture can deepen the conditions under which good care breaks down, but its connective infrastructure can also provide the structural support that makes care sustainable. Recent work in HCI has called for designing not only to fix, protect, or prevent, but to actively support human flourishing~\cite{positech2024, positech2025, to2023flourishing, kim2026socialmediafeellike}. The two prior iterations of this workshop series~\cite{positech2024, positech2025} explored how social technologies might be designed from the ground up to support positive social outcomes. This reorientation is especially relevant here, where the question is not just how to reduce harm but how to build the conditions for sustained collective care.

\subsubsection{How Platform Architecture Undermines the Caring Process}
Media exposure to large-scale crises---including terrorism, mass violence, and public health emergencies---amplifies distress and prolongs psychological impact \cite{holman2014media, garfin2018aftermath, garfin2020novel, thompson2019media}. Repeated exposure to emotionally salient content has long been associated with cycles of stress, worry, and withdrawal in broadcast media environments.

What distinguishes contemporary social media is not the existence of this dynamic, but its scale, persistence, and structural embedding in platform design:

\begin{itemize}
    \item Platforms expose users to suffering from everywhere, continuously, with no natural boundary. Unlike a newspaper that ends or a broadcast that finishes, feeds are infinite. As a result, the responsibility for setting boundaries falls entirely on the individual. This individualization of boundary-setting transforms attentiveness from a situational response into an ongoing self-regulation problem. Meanwhile, the actions available remain the same: like, share, comment, scroll. This mismatch between boundless awareness and limited agency is precisely the condition under which empathic distress, rather than compassion, takes hold~\cite{singer2014empathy}.
    \item Algorithmic curation compounds this. Platforms optimize for engagement, and emotionally intense content drives engagement more than nuanced or solution-oriented content. Users trying to stay informed encounter a systematically skewed picture that pushes toward emotional dysregulation rather than constructive response~\cite{kim2025privacy}.
    \item The dominant platforms among young people are structured around broadcast rather than participation. Influencers address large audiences; regular users consume, react, and share. Such asymmetric dynamics reinforce the gap between awareness and action: users witness problems but have few routes to participate in addressing them.
    This pattern reflects a broader shift in social media ecosystems as they scale and commercialize. Platforms that begin as more participatory spaces often evolve toward broadcast dynamics, concentrating visibility and influence among a small subset of users. In Tronto's terms, this structure leaves responsibility diffuse: many users are exposed to the same problems, but few are positioned to take meaningful action.
    \item This architecture also cultivates visible expressions of concern in place of material response. Posting about an issue can create an illusion of meaningful impact that displaces rather than motivates sustained commitment~\cite{morozov2009brave}, a pattern described as ``slacktivism''~\cite{kristofferson2014nature}. Experimental evidence supports this: when initial acts of support are publicly visible, as they typically are on social media, people are less likely to engage in subsequent meaningful action~\cite{kristofferson2014nature}. Social media lowers the cost of expressing outrage while decoupling that expression from the prosocial behavior it ordinarily accompanies~\cite{crockett2017moral}, producing a net increase in anxiety and moral reproach without a corresponding increase in care.
\end{itemize}

\subsubsection{How Platform Infrastructure Could Support Sustainable Care}
The same platforms that stall care at its earliest phase have demonstrated capacity to support later phases under the right conditions~\cite{Briggs2024GenPresentc}, though this potential remains unevenly realized.

\begin{itemize}
    \item At the level of \textit{responsibility} (phase 2), mutual aid networks and crowdfunding campaigns show how platforms can help specific people assume responsibility for specific needs. Research on movements from the Arab Spring to Black Lives Matter documents how platforms facilitate rapid organization and the scaling of local efforts into broader coalitions~\cite{tufekci2017twitter, jackson2020hashtagactivism}. These examples work because they distribute the caring burden rather than leaving it ambient.
    \item At the level of \textit{competence} (phase 3), framing matters: research on climate communication finds that maintaining hope and a sense of agency sustains engagement, while helplessness discourages it~\cite{ojala2012hope}. Studies in rural settings also highlight that mobile technologies need to reinstate the potentials they afford, which strengthens sustained adoption and effective use \cite{wyche2016don}. When platforms surface what people can actually do, not just what is going wrong, they begin to support the knowledge and capacity that competent caregiving requires.
    \item At the level of \textit{responsiveness} (phase 4), online communities enable people to process difficult experiences together. Research on mental health communities documents how platforms facilitate emotional support~\cite{andalibi2017sensitive}.
\end{itemize}

When platforms support shared ownership of the caring process and ongoing conversation about what is needed, they approach Tronto's fifth phase, \textit{caring with}, and create the conditions under which good care becomes sustainable. The design question is whether such conditions can be supported architecturally, rather than emerging only in spite of platform design.

These examples, however, share a common assumption: that the populations in question are already online and embedded in platform ecosystems. Studies conducted in the Global South and rural settings reveal stark disparities not only in connectivity but in the institutional, educational, and cultural conditions that shape whether digital tools can support collective action at all \cite{hardy2019rural}. In many of these contexts, the caring process unfolds through non-digital channels that platform-centric design frameworks tend to overlook. Designing for sustainable care without addressing these asymmetries risks producing solutions that serve already-connected populations while leaving out communities most affected by climate change, extractive economies, and humanitarian crises.

\subsection{Relevance to the CSCW Community}
This workshop continues the \href{https://positech.github.io/}{Positech community}'s workshop series (\href{https://positech-cscw-2024.github.io/}{CSCW 2024}, \href{https://positech-cscw-2025.github.io/}{2025})~\cite{positech2024, positech2025}, which has examined how social technologies might support human flourishing. The majority of the current organizers participated in or organized previous iterations, and the community has grown each year. In this sense, the workshop is itself an exercise in what it studies: a group of researchers sustaining collective engagement with a difficult problem across years, building shared responsibility through repeated collaboration.
\section{Workshop Themes}
\label{sec:themes}

This workshop asks what it would take to design social technologies that support ``sustainable care''~\cite{kim2026sustainable}. Using Tronto's Ethics of Care framework~\cite{tronto1993moral, tronto2013caring} as the organizing structure, we invite participants to examine how platform features stall or advance each phase of the caring process. As laid out in Section~\ref{sec:introduction}, sustainable care requires communities that share responsibility. This workshop accordingly foregrounds collective action as a central concern, asking how platform design can support not just individual well-being but the organized, shared engagement that sustained care demands. The themes below describe phases of a single process rather than independent problems, and they serve as the through-line for every workshop activity.

These themes describe phases of a single process rather than independent problems, and they serve as the through-line for every workshop activity.

\begin{enumerate}
    \item \textbf{Attentiveness without overwhelm} (\textit{caring about}). Platforms flood users with awareness that has no natural endpoint. What would ``enough for today'' look like as a design pattern, and what algorithmic alternatives might balance awareness of problems with awareness of responses?

    \item \textbf{Distributing responsibility} (\textit{caring for}). Broadcast architecture leaves responsibility diffuse: content is shared widely, but no one is positioned to act. How might platforms connect users to communities where responsibility is shared and made specific?

    \item \textbf{Supporting competent action} (\textit{caregiving}). Current platforms privilege emotional framing over actionable knowledge. How might concrete, achievable actions be surfaced alongside awareness of problems? What design patterns support organized, effective collective response?

    \item \textbf{Making responsiveness visible} (\textit{care receiving}). Negative stories dominate, with little visibility into whether care efforts are working. How might design help users see collective progress, process difficult information into actionable concern, and attend to what their efforts have and have not accomplished?

    \item \textbf{Building solidarity} (\textit{caring with}). Current platforms present users as isolated witnesses to collective problems. How might design foster communities where caregivers and care receivers alike participate in defining needs and deciding how to meet them?
\end{enumerate}

\section{Workshop Logistics}
The workshop is a full-day, in-person session structured around three collaborative activities, a panel discussion, and collective reflection (see Table~\ref{tab:schedule} for full schedule). Participants will be grouped by domain of interest (e.g., climate, public health, civic engagement), surveyed before the workshop, so that each activity builds cumulatively: from understanding what good care looks like in a given domain, to diagnosing how social media currently helps or hinders that care, to generating concrete design directions. The day closes with large-group reflection on research questions for the CSCW community around sustainable care.

We will accept 20--25 participants (excluding organizers), recruited through social media, mailing lists, and the workshop website. Participants will be selected based on relevance of their research or practice to the workshop themes and diversity of disciplinary perspective. The workshop requires a room with a projector, microphone, and movable seating for small-group work.

\begin{table*}[!ht]
\centering
\caption{Workshop schedule. This full-day, in-person workshop brings together researchers and designers to examine how social technology can support sustainable care for people engaging with societal challenges. Each activity returns to themes in Section~\ref{sec:themes} as a lens for analyzing current platforms, generating design directions, and identifying open research questions. (Workshop Website: \href{https://positech.github.io/cscw2026.html}{\protect\textcolor{teal}{https://positech.github.io/cscw2026.html}})}
\label{tab:schedule}
\begin{tabular}{@{}p{1.8cm}p{3.8cm}p{8.8cm}@{}}
\toprule
\textbf{Time} & \textbf{Activity} & \textbf{Description} \\
\midrule
09:00--09:25 & Welcome \& Introductions & Agenda, goals, and participant introductions \\
09:25--09:45 & Framing Sustainable Care & Introduction to ethics of care framework and the five workshop themes \\
\midrule
09:45--10:30 & \textbf{Activity 1:} Domain-Specific ``Good Care'' & Small groups (by domain): What does good care look like in domain \textit{X}? Identify cases of good care across all five phases of the caring process \\
10:30--10:50 & \textit{Break} & \\
10:50--11:15 & Report-Back & Each group shares key patterns \\
\midrule
11:15--12:00 & \textbf{Activity 2:} Social Technology \& Care & New groups (by technology): How is social technology currently helping or hindering care in domain \textit{X}? \\
12:00--12:25 & Report-Back & Each group shares findings \\
12:25--14:05 & \textit{Lunch} & \\
14:05--14:30 & Rest \& Reflect & Nap time, stretch, and individual reflection \\
\midrule
14:30--15:15 & \textbf{Activity 3:} Design Directions & Return to domain groups: How might social technology be designed differently? What concrete changes are feasible, and what remains wicked? \\
15:15--15:45 & \textit{Break} & \\
\midrule
15:45--16:30 & Large-Group Reflection & Sharing across groups; identifying key research questions for the CSCW community on sustainable care \\
16:30--17:00 & Closing & Feedback on the workshop, community building plan, and next steps \\
\bottomrule
\end{tabular}
\end{table*}

\section{Organizers}

*All authors of this proposal have committed to attending the workshop in person.

\noindent\textbf{JaeWon Kim} is a PhD candidate at the University of Washington Information School. Her research focuses on understanding, designing, and building social technologies that center on meaningful social connections, especially for the youth.

\textbf{Lindsay Popowski} is a PhD candidate at Stanford HCI. She designs and builds social computing systems that re-imagine social media's fundamental design choices in order to create online experiences that help, rather than hinder, well-being.

\noindent\textbf{Louisa Conwill} is a PhD candidate in computer science and engineering at the University of Notre Dame. She researches how religious and philosophical traditions can inspire more positive visions for technology design.

\noindent\textbf{Elizabeth 'Lizzie' Li} is a PhD candidate in media, technology, and society at Northwestern University. Her research focuses on how to empower social media users to to navigate social media in ways that foster authenticity and wellbeing.

\noindent\textbf{Meryl Ye} is a PhD student at Carnegie Mellon University. She studies how sociotechnical systems shape human judgment, autonomy, and wellbeing.

\noindent\textbf{Jiaying `Lizzy' Liu} is an Assistant Professor at the Nanyang Technological University. She studies the societal and ethical dimensions of multimodal and AI-mediated communication, such as video-sharing platforms and AI companions.

\noindent\textbf{Jose A. Guridi} is a PhD candidate in Information Science at Cornell University. His research examines AI governance across policy, organizational, and design levels, drawing on cross-country fieldwork and HCI approaches to responsible design, adoption, and use of AI.

\noindent\textbf{Theia Henderson} is a fifth-year PhD student in computer science at MIT CSAIL. Her research focuses on improving social media and online social tools, building real-world systems informed by her background in theoretical computer science.

\noindent\textbf{Bingxu Han}, PhD at Stanford University, studies AI communication and humanistic psychotherapy. She uses experiments to test how LLMs can facilitate communication between care-providers and support-seekers in healthcare and mental healthcare contexts.    

\noindent\textbf{Dennis Wang} is an Assistant Professor of Computer Science at National Yang Ming Chiao Tung University, Taiwan. He designs, builds, and evaluates social computing systems for social connectedness, focusing on how online norms emerge and can be shaped positively.

\noindent\textbf{Angel Hwang} is an Assistant Professor of Communication and Computer Science at the University of Southern California. Her work examines the impact of AI applications on users' interpersonal relationships (in both professional and personal settings), psychological well-being, and mental health. 

\noindent\textbf{Susan Wyche} is an Associate Professor at Michigan State University. She studies technology use in rural Kenya. Her findings draw attention to barriers (cost, access, and digital literacy) while identifying inclusive design opportunities.

\noindent\textbf{Yasmine Kotturi}
Yasmine Kotturi is an Assistant Professor of Human-Centered Computing at UMBC, directing a lab on community-engaged AI and worker resilience. She builds and deploys social technologies with underserved communities. She is a CRA Trustworthy AI and FASPE Fellow.

\noindent\textbf{Gillian Hayes} is Kleist Professor of Informatics at UCI. Her community-based research focuses on assistive, health, and learning technologies. She is the co-director of CERES, a global research network around youth and technology.

\noindent\textbf{Angela D. R. Smith} is an Assistant Professor at the University of Texas at Austin's School of Information and co-founder of the REALITY (Research on Equity, Access, and incLusion in Technology and societY) Lab. A critical scholar whose work draws on race-based methodologies, she examines how historically marginalized communities create, resist, and imagine through digital technologies, centering care, abundance, and community flourishing.

\begin{acks}
JaeWon Kim would like to acknowledge the CERES Network, University of Washington Global Innovation Funds (GIF), and Student Technology Funds (STF), which provided support for this work. This work was also funded in part by the Paul G. Allen School of Computer Science \& Engineering Endowed Fund for Excellence and a gift from Google.
\end{acks}

\bibliographystyle{ACM-Reference-Format}
\bibliography{newrefs}

\end{document}